%% file: main.tex
\documentclass[]{style/ceurart}
\usepackage{listings}
\usepackage{tikz}
\usetikzlibrary{positioning,arrows.meta,fit}

\begin{document}

\copyrightyear{2026}
\copyrightclause{Copyright for this paper by its authors.
  Use permitted under Creative Commons License Attribution 4.0
  International (CC BY 4.0).}

\conference{CLEF 2026: Conference and Labs of the Evaluation Forum, September 21-24, 2026, Jena, Germany}

\title{Can Tokens Compete? Token Representations against Supervised CNN Backbones for BirdCLEF+ 2026}

\author[1]{Anthony Miyaguchi}[
    orcid=0000-0002-9165-8718,
    email=acmiyaguchi@gatech.edu,
]
\cormark[1]

\author[1]{Murilo Gustineli}[
    orcid=0009-0003-9818-496X,
    email=murilogustineli@gatech.edu,
    url=https://murilogustineli.com,
]
\cormark[1]

\author[1]{Adrian Cheung}[
    orcid=0009-0006-8650-4550,
    email=acheung@gatech.edu,
]
\cormark[1]

\address[1]{Georgia Institute of Technology, 225 North Ave NW, Atlanta, GA 30332}
\cortext[1]{Corresponding author.}

\begin{abstract}
This paper details the DS@GT ARC team's approach to BirdCLEF+ 2026, multi-label detection of animal vocalizations in soundscapes from the Pantanal wetlands.
The 2026 edition adds about an hour of labeled soundscapes, shifting the task toward supervised pipelines fit to the labeled set.
First, we build a competitive supervised baseline that ensembles a frozen Perch v2 backbone, a trained HGNetV2-B0 sound-event-detection network, and a non-bird prototypical head, reaching a private leaderboard score of 0.936 at rank 1894 within a 90-minute CPU budget.
Second, we ask whether token-based representations can compete, contrasting codec representations from neural audio codecs against semantic representations from foundational embeddings.
We compare two bioacoustic specialist models against four token-based encoders trained on AudioSet.
The repository for this work can be found at \url{https://github.com/dsgt-arc/birdclef-2026}.
\end{abstract}


\begin{keywords}
  BirdCLEF \sep
  Bioacoustics \sep
  Soundscape classification \sep
  Audio foundation models \sep
  Neural audio codecs
\end{keywords}

\maketitle

\input{sections/00_main}

\bibliography{main}

\clearpage
\appendix
\input{sections/90_appendix_baseline}

\input{sections/90_appendix_bands}

\input{sections/90_appendix_perch}
\input{sections/90_appendix_encoders}

\input{sections/90_appendix_ssamba}

\end{document}

%% file: sections/00_main.tex

\input{sections/10_introduction}
\input{sections/20_related_work}
\input{sections/30_data}

\input{sections/40_baseline}            
\input{sections/50_token_representations} 
\input{sections/60_discussion}
\input{sections/70_closing}             

%% file: sections/10_introduction.tex
\section{Introduction}

BirdCLEF+ \cite{birdclef2026overview} is a competition, hosted on Kaggle and organized through the LifeCLEF lab at the Conference and Labs of the Evaluation Forum (CLEF) \cite{lifeclef2026}, to build a model that detects and classifies animal vocalizations in soundscapes.
Soundscapes are recordings of a landscape captured by autonomous recording units deployed to the field.
The 2026 edition of the competition is focused on multi-label detection on 60-second soundscapes from the Pantanal wetlands in South America.

The largest challenge in the previous years was that while there were many recordings for each species, they were weakly labeled.
A given audio clip in the focal set contains vocalizations for a particular species, but the time range in which the vocalization occurs is not given, and there may also be multiple sources of sound.
The focal recordings are also biased toward the most common species that can be found via population centers, because they are easier to capture and there are likely to be more people around to capture those sounds.
This heavy-tailed distribution makes species with only a handful of examples hard to learn, and the problem formulation encourages few-shot solutions.

The 2026 edition of the competition added about an hour of labeled soundscapes, leading to a significant change in the strategy for model development.
Previous editions focused on the domain shift from focal recordings taken from Xeno-canto \cite{xenocanto}, which are crowd-sourced, to the soundscape domain, which is multi-label and without provided ground truth.
This means that supervised algorithms now dominate the leaderboard and lean heavily into the small set of labeled data.

First, we build a supervised baseline system following the best practices established by the Kaggle community, leading to a competitive leaderboard submission on the labeled soundscape set for team DS@GT ARC at rank 1894 with a score of 0.936.
Second, we investigate whether token-based representations can be effective on this leaderboard, and compare the semantic representations produced by foundational audio embeddings against the reconstruction-oriented codec representations produced by neural audio codecs.
We close with discussion around a future of foundational systems capable of contextualizing soundscapes, beyond the domain of short focal recordings.

%% file: sections/20_related_work.tex
\section{Related Work}

Soundscape recognition typically uses deep networks on audio spectrograms.
The usual representation is the mel-scaled short-time Fourier transform (STFT), which scales frequencies to the perceptual scale of human hearing and is fed into convolutional networks (CNNs) whose spatial bias suits the spectral structure of sound.
BirdNET \cite{kahl2021birdnet} and Perch v2 \cite{hamer2025perch2} are the state of the art, training CNNs to produce both windowed species occurrence probabilities and a reusable embedding space.
Global birdsong embeddings demonstrate a large domain gap in transfer learning with respect to general-purpose foundation models like AudioMAE for bioacoustics \cite{ghani2023global}.
Self-supervised encoders like AVES \cite{hagiwara2022aves}, benchmarks like BirdSet \cite{rauch2025birdset}, and a recent comparative review \cite{schwinger2025foundationreview} help contextualize efforts toward exploiting larger bioacoustic datasets.

We study two families of token representations.
Semantic embeddings come from self-supervised audio models: AST \cite{gong2021ast} adapted vision transformers to spectrograms, BEATs \cite{chen2023beats} and EAT \cite{chen2024eat} improved this with masked prediction, wav2vec~2.0 \cite{baevski2020wav2vec} learns from speech, and Bird-MAE \cite{rauch2025birdmae} adapts masked autoencoding to birds.
Codec representations come from neural audio codecs: SoundStream \cite{zeghidour2021soundstream}, EnCodec \cite{defossez2022encodec}, and the single-codebook WavTokenizer \cite{ji2024wavtokenizer} are trained to reconstruct waveforms for human perception.
We also test linear-time state-space models, where Mamba \cite{gu2023mamba}, SSAMBA \cite{shams2024ssamba}, and recent bioacoustic SSMs \cite{tang2025ssmbioacoustics} scale to the long sequences needed to contextualize a full soundscape.

The recurring difficulty in earlier editions was domain shift: focal recordings from Xeno-canto \cite{xenocanto} had to generalize to unlabeled multi-label soundscapes \cite{ghani2024domaingap}, and the 2026 labeled hour shifts this toward supervised pipelines fit to the labeled set.
The DS@GT team has competed since 2022: motif mining \cite{miyaguchi2022motif} in 2022 \cite{kahl2022birdclefoverview}, transfer learning on BirdNET embeddings \cite{miyaguchi2023transfer} in 2023 \cite{kahl2023birdclefoverview}, sound-separation pseudo-labels \cite{miyaguchi2024pseudomultilabel, denton2021improvingbirdclassificationunsupervised} in 2024 \cite{kahl2024birdclefoverview}, and a word2vec-style token model \cite{miyaguchi2025distilling, mikolov2013word2vec} in 2025 \cite{canas2025birdclefoverview}.
We build upon these techniques, reusing the vectorized knowledge of models in sample-efficient ways.

%% file: sections/30_data.tex
\section{Data}

BirdCLEF 2026 data contains 522.1 hours of audio, split between 344.5\,h of focal recordings across 206 species and 177.6\,h of soundscapes from 23 sites in the Pantanal, Brazil.
Only 1.03\,h (66 files, 739 unique 5\,s windows, spanning 9 of the 23 sites) of the soundscape audio is labeled; the remaining 176.5\,h is unlabeled.
Of the 234 target species, 28 are derived from soundscapes alone, many of those being insects.
Only 2.4\% of focal recordings come from the Pantanal bounding box, and 87 species have zero recordings from Pantanal.

\input{sections/floats/40_fig_taxa-spectrograms}

\input{sections/floats/30_tab_dataset-asymmetry}

The animal vocalizations are not necessarily separable by frequency band.
Frogs, birds, and insects all concentrate energy in 1--4\,kHz and differ in spectro-temporal structure (Figure~\ref{fig:taxa-spectrograms}).

A statistical breakdown of the training and test data (Table~\ref{tab:dataset-asymmetry}) reiterates the challenges posed by the competition and lab.
The gap between recording domains, together with the long-tailed distribution of recordings over species, means that models must be sample-efficient and account for the low-resource nature of the data.

%% file: sections/floats/40_fig_taxa-spectrograms.tex
\begin{figure}[ht]
  \centering
  \includegraphics[width=\textwidth]{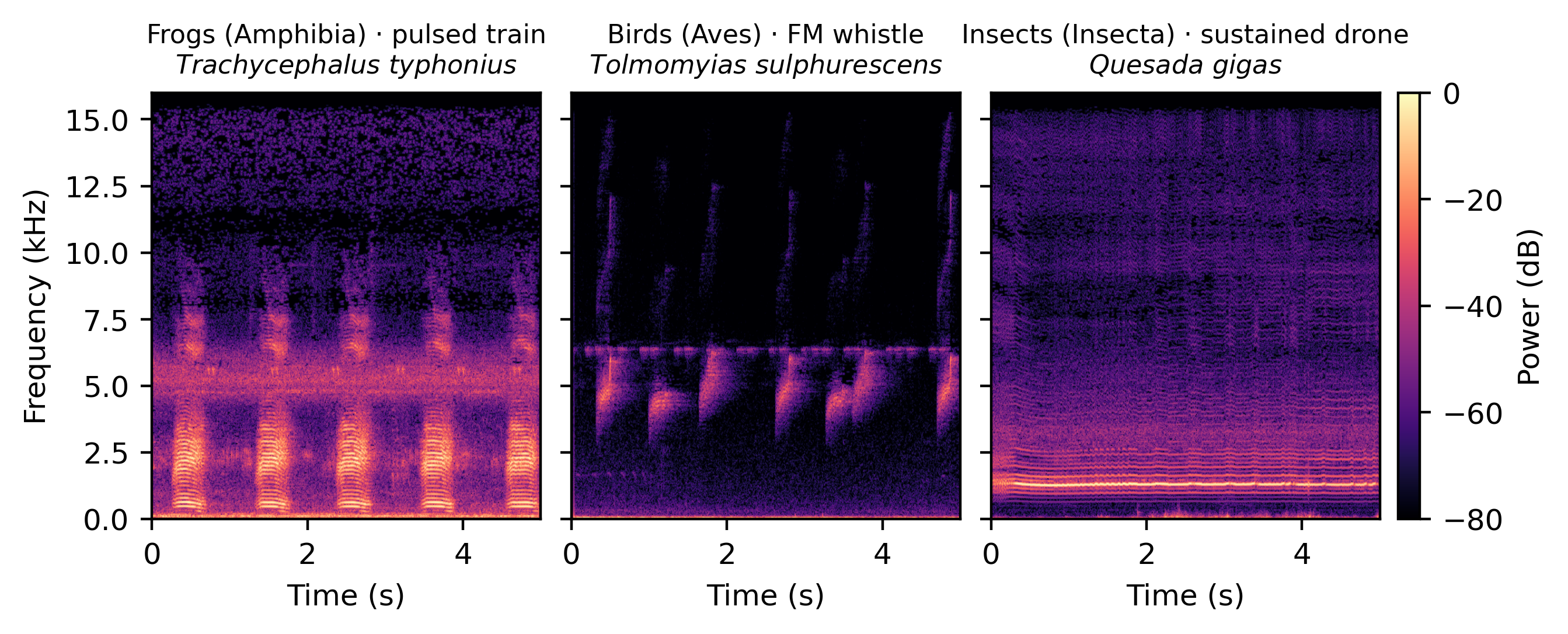}
  \caption{
    Representative spectrograms of three vocalizing taxa in BirdCLEF 2026.
    The taxa overlap in frequency.
    Frogs produce pulsed call trains, birds frequency-modulated whistles, and insects a sustained broadband droning sound.
  }
  \label{fig:taxa-spectrograms}
\end{figure}

%% file: sections/floats/30_tab_dataset-asymmetry.tex
\begin{table}[h]
  \centering
  \caption{
  Class composition of the focal training set vs. the labeled soundscape windows.
  The focal training is mostly birds, while the soundscape is mostly amphibians.
  In the insect class, the focal species do not overlap with the soundscape species.
  Note that soundscape windows are multi-label, and thus the percentage of windows sums up to a value greater than 100.
    }
  \label{tab:dataset-asymmetry}
  \begin{tabular}{lrrrr}
    \toprule
             & \multicolumn{2}{c}{Focal (recordings)} & \multicolumn{2}{c}{Soundscape
                                                          (windows)}                         \\
    \cmidrule(lr){2-3} \cmidrule(lr){4-5}
    Class    & Species                                & \% recs                       & Species & \% windows \\
    \midrule
    Aves     & 162                                    & 97.89                         & 28      & 28.7       \\
    Amphibia & 32                                     & 1.27                          & 17      & 76.6       \\
    Insecta  & 3                                      & 0.56                          & 25      & 22.7       \\
    Mammalia & 8                                      & 0.28                          & 4       & 5.5        \\
    Reptilia & 1                                      & 0.003                         & 1       & 1.8        \\
    \bottomrule
  \end{tabular}
\end{table}

%% file: sections/40_baseline.tex
\section{Supervised Baseline System}
\label{sec:baseline}


The 2026 data rewards supervised pipelines fit directly to the hour of labeled soundscape data, and we follow Kaggle community best practices to build a competitive baseline.
The system is an ensemble of a frozen Perch v2 \cite{hamer2025perch2} backbone with MLP probes whose probabilities are blended with a trained HGNetV2-B0 sound-event-detection (SED) network.
The best model, which includes a prototypical head for non-bird species, reaches a public leaderboard score of \textbf{0.931} and a private score of \textbf{0.936} at rank 1894 (Figure~\ref{fig:baseline-ensemble}, Table~\ref{tab:baseline-ablation}).
Most score gains come from frontend and architecture choices rather than training tricks or extra data, and we've aimed for a parsimonious system.

\subsection{Methodology}

\input{sections/floats/40_fig_ensemble}

The ensemble uses two different backbones.
The first is Perch v2 \cite{hamer2025perch2} which is frozen with no backbone training.
It emits a 1536-dimensional embedding and 14{,}795-class logits per 5\,s window, which we map onto the 234 competition species (203 direct scientific-name matches, 6 genus proxies, 25 insect sonotypes suppressed).
The second is an HGNetV2-B0 convolutional SED network operating on log-mel spectrograms (256 mel bins, 32\,kHz, 2048-point FFT, 313-sample hop).

For the Perch member we fit logit-space MLP probes on the frozen embeddings over the 59 fully-labeled soundscapes (of the 66 labeled files) using 5-fold out-of-fold (OOF) GroupKFold cross-validation.
We also explored ProtoSSM, a temporal head that refines the per-window probe logits across the twelve 5\,s windows of a soundscape.
It uses a bidirectional state-space model and a per-species prototype layer \cite{snell2017prototypical} with a learned gate against probe logits.
The SED member trains HGNetV2-B0 on the focal and soundscape recordings across 4 folds, with spectrogram mixup and a multi-label binary cross-entropy loss.
The SED member adds dB-scale mel normalization (AmplitudeToDB) and higher mel resolution.

At inference each 60\,s soundscape is split into 12 non-overlapping 5\,s windows, and each member emits a 234-species probability vector per window.
The two members are averaged together with equal weight ($w_{\text{Perch}}=0.5$), with the SED member additionally test-time augmented across its 4 folds.
Site- and hour-conditioned priors are then applied as post-processing.

\paragraph{Non-Bird Prototypical Head}
Of the 234 competition species, 72 are non-bird taxa outside the training labels of Perch. 
To score these species, we introduce a prototypical head \cite{snell2017prototypical} that uses the Perch backbone.
Although Perch was trained exclusively on avian audio, it captures generalized acoustic features that help cluster non-bird vocalizations.
Of the 72 non-bird taxa, 47 appear in the labeled soundscapes, allowing us to construct positive ($\boldsymbol{\mu}^+_s$) and negative ($\boldsymbol{\mu}^-_s$) prototypes by averaging the Perch embeddings of all windows where the species is present, and all windows where it is absent, respectively.
We compute a cosine-difference score for each species $s$ and window $t$:
\begin{align*}
  \hat{y}^{\text{proto}}_{t,s} = \text{clip}(\cos(\mathbf{e}_t, \boldsymbol{\mu}^+_s) - \cos(\mathbf{e}_t, \boldsymbol{\mu}^-_s),\;0,\;1)
\end{align*}
For all non-bird taxa, we first take the SED predictions directly.
If a prototype is available for a species, these precomputed scores are merged via element-wise maximum with the SED output:
$\hat{y}^{\text{final}}_{t,s} = \max\left(\hat{y}^{\text{sed}}_{t,s},\,\hat{y}^{\text{proto}}_{t,s}\right)$.
Because the prototype matrices reuse Perch embeddings from the bird head, they add little inference overhead.
The full pipeline completes in roughly 64 minutes, well within the 90-minute CPU limit.

\subsection{Leaderboard Results}

The ensemble performs better than either of its constituents.
The frozen Perch probe scores 0.920 and the SED member 0.917, while their blend reaches 0.929 (Table~\ref{tab:baseline-ablation}).
The simplest Perch member uses a ridge regression probe \cite{hoerl1970ridge} scoring 0.866 and is replicated in later experiments such as that in Table~\ref{tab:birdclef2026-loso}.
Dropping augmentations like the ProtoSSM and the site/hour post-processing leads to only marginal degradation in leaderboard performance.

Adding the non-bird prototypical head to the MLP-only ensemble raises the public score to 0.931 and the private score to 0.936.
Offline evaluation on the 66 labeled soundscapes shows the prototype substantially outperforms the SED model alone for non-bird taxa (macro AP 0.895 vs. 0.074).
This confirms that the Perch embedding space carries useful information for non-bird species, despite only being trained on bird data.

\input{sections/floats/40_tab_baseline-ablation}

\input{sections/floats/40_tab_sed-ladder}

Table~\ref{tab:sed-ladder} contains an ablation of architectural changes to the SED branch of the ensemble.
Frontend and backbone choices contribute to most of the score gains.
dB-scale mel normalization is the largest frontend lever by making quiet and loud recordings comparable, which matters when focal training clips and soundscape test clips differ sharply in level.
We try a few more elaborate choices like an ImageNet EfficientNet-B0 SED distilled from Perch v2 embeddings which recovers an identical single fold, and an EfficientNetV2-S pretrained on tropical Xeno-Canto audio (Appendix~\ref{sec:appendix-baseline}).
Neither buys significant accuracy, and both add complexity through teacher pipelines or large-scale training.

Offline validation AUROC anti-correlated with the public leaderboard across 20+ submissions, so we treated every change as a single-variable experiment.
A curated record is in Appendix~\ref{sec:appendix-baseline} (Table~\ref{tab:baseline-negatives}).

%% file: sections/floats/40_fig_ensemble.tex
\definecolor{lbgreen}{HTML}{157347}
\providecommand{\lbscore}[1]{{\itshape\color{lbgreen}#1}}
\providecommand{\dt}[1]{{\tiny\color{black!60}#1}}
\begin{figure}[b]
  \centering
  \begin{tikzpicture}[
      node distance=4mm and 6mm,
      font=\scriptsize,
      box/.style={draw, rounded corners=2pt, align=center, inner sep=3pt, minimum height=9mm},
      perch/.style={box, fill=SkyBlue!25},
      sed/.style={box, fill=Salmon!25},
      proto/.style={box, fill=ForestGreen!15},
      merge/.style={box, fill=Gray!18},
      io/.style={box, fill=white},
      grp/.style={draw=black!35, dashed, rounded corners=3pt, inner sep=2.4mm},
      glbl/.style={font=\tiny\itshape, inner sep=1pt},
      arr/.style={-{Latex[length=1.8mm]}, semithick},
    ]
    \node[io] (win) {60\,s soundscape\\\dt{$\to$ 12 $\times$ 5\,s windows}};

    \node[perch, above right=4mm and 5mm of win] (perch)
      {Perch~v2 \dt{(frozen)}\\\dt{1536-d embedding}};
    \node[perch, right=of perch] (mlp)
      {MLP probes\\\dt{logit-space, 5-fold OOF}\\\lbscore{0.920}};

    \node[sed, below right=4mm and 5mm of win] (hgnet)
      {HGNetV2-B0 SED\\\dt{log-mel, mixup, BCE}};
    \node[sed, right=of hgnet] (tta)
      {4-fold $+$ TTA\\\lbscore{0.917}};

    \node[merge, right=5mm of mlp.east |- win] (blend)
      {prob-space blend\\\dt{$w_{\text{Perch}}{=}0.5$}\\\lbscore{0.922}};
    \node[merge, right=of blend] (prior)
      {site/hour\\priors\\\lbscore{$+0.007$}};
    \node[merge, right=of prior] (nbmerge)
      {non-bird\\merge\\\lbscore{$+0.002$}};
    \node[io, right=of nbmerge] (sub)
      {submission\\\dt{234 probs}\\\lbscore{\textbf{0.931}}};

    \node[proto, below=7mm of nbmerge, minimum height=0pt, inner sep=2pt] (proto)
      {non-bird\\prototypes};

    \node[grp, fit=(perch)(mlp)] (grpA) {};
    \node[grp, fit=(hgnet)(tta)] (grpB) {};
    \node[glbl, text=SkyBlue!45!black, above=0.2mm of grpA.north west, anchor=south west]
      {Member 1: frozen, probes only};
    \node[glbl, text=Salmon!45!black, below=0.2mm of grpB.south west, anchor=north west]
      {Member 2: trained on audio};

    \draw[arr] (win) |- (perch);
    \draw[arr] (win) |- (hgnet);
    \draw[arr] (perch) -- (mlp);
    \draw[arr] (hgnet) -- (tta);
    \draw[arr] (mlp) -| (blend);
    \draw[arr] (tta) -| (blend);
    \draw[arr] (blend) -- (prior);
    \draw[arr] (prior) -- (nbmerge);
    \draw[arr] (proto) -- (nbmerge);
    \draw[arr] (nbmerge) -- (sub);
  \end{tikzpicture}
  \caption{
    The supervised baseline is an equal-weight, two-member ensemble extended with a non-bird prototypical head.
    Each 60\,s soundscape is split into 12 non-overlapping 5\,s windows that feed both members: a \emph{frozen} Perch~v2 backbone read through logit-space MLP probes, and a trained HGNetV2-B0 SED network on log-mel spectrograms.
    Their 234-species probability vectors are blended equally, post-processed with site/hour priors, and merged with non-bird prototype scores.
    The whole pipeline runs in $\sim$64\,min, within the 90\,min CPU limit.
    Public LB scores are shown in green.}
  \label{fig:baseline-ensemble}
\end{figure}
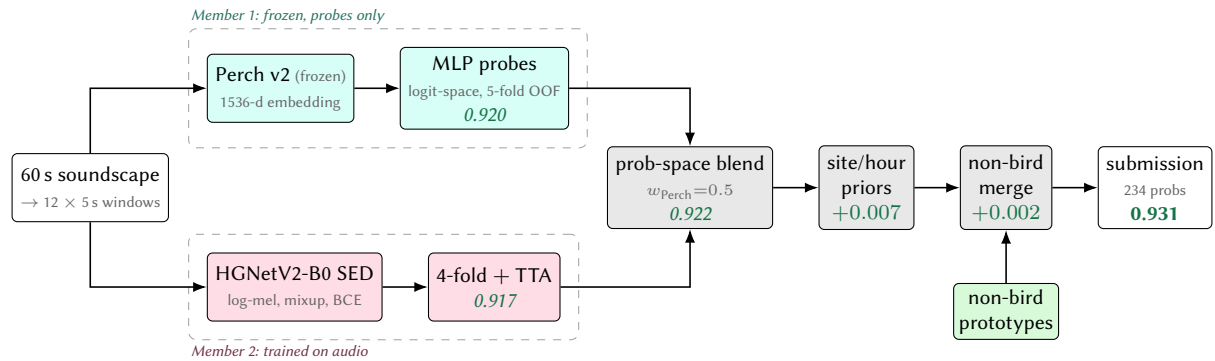

%% file: sections/floats/40_tab_baseline-ablation.tex
\begin{table}[h]
  \centering
  \caption{
    Baseline ablations on the BirdCLEF 2026 Kaggle leaderboard.
    $\Delta$ is the public-LB gain over the frozen ridge probe.
  }
  \label{tab:baseline-ablation}
  \begin{tabular}{lrrr}
    \toprule
                  & \multicolumn{2}{c}{Kaggle LB} &                \\
    \cmidrule(lr){2-3}
    Configuration & pub            & priv         & $\Delta$       \\
    \midrule
    \multicolumn{4}{l}{\emph{Members (standalone)}} \\
    Perch v2 + ridge probe (frozen)   & 0.866          & 0.862          & ---      \\
    Perch v2 + MLP probes (frozen)    & 0.920          & 0.918          & $+0.054$ \\
    HGNetV2-B0 SED (4-fold)           & 0.917          & 0.924          & $+0.051$ \\
    \midrule
    \multicolumn{4}{l}{\emph{Ensemble} ($w_{\text{Perch}}=0.5$)} \\
    \quad with ProtoSSM head          & 0.927          & 0.923          & $+0.061$ \\
    \quad MLP-only (drop ProtoSSM)    & 0.929          & 0.925          & $+0.063$ \\
    \quad no post-processing          & 0.922          & 0.916          & $+0.056$ \\
    \multicolumn{4}{l}{\emph{Non-bird prototypical head}} \\
    \quad MLP-only + non-bird head    & \textbf{0.931} & \textbf{0.936} & $+0.065$ \\
    \bottomrule
  \end{tabular}
\end{table}

%% file: sections/floats/40_tab_sed-ladder.tex
\begin{table}[h]
  \centering
  \caption{
    Architectural ablations from the SED member.
    Each row is the best public-LB checkpoint at that stage.
    $\Delta$ is relative to the row above.
  }
  \label{tab:sed-ladder}
  \begin{tabular}{llrr}
    \toprule
    Stage              & Change                                   & pub             & $\Delta$       \\
    \midrule
    Base SED           & EfficientNet-B0 SED ($+$ post-proc.)     & 0.827          & ---            \\
    $+$ dB-mel         & high-res mel $+$ AmplitudeToDB           & 0.859          & $+0.032$       \\
    $+$ HGNetV2-B0     & backbone swap $+$ NNAudio dB-mel         & 0.906          & $+0.047$       \\
    $+$ 4-fold         & logit-mean over folds                    & 0.913          & $+0.007$       \\
    $+$ priors         & site/hour post-processing                & 0.917          & $+0.004$       \\
    \midrule
    $+$ Perch probe    & score-level blend (ensemble)             & \textbf{0.929} & $+0.012$       \\
    \bottomrule
  \end{tabular}
\end{table}

%% file: sections/50_token_representations.tex
\section{Token Representations for Bioacoustic Soundscapes}

Having established a supervised baseline, we ask whether token-based representations can compete with it on this leaderboard.
We compare codec representations from neural audio codecs, which are trained to reconstruct the waveform, versus semantic representations from foundational audio embeddings, which are trained to preserve similarity and separability.
We rule out the reconstruction route, then turn to the semantic embeddings for their versatility.

\subsection{Codec Representations}

We tested whether a discrete neural audio codec could replace embeddings for species retrieval.
We used WavTokenizer-large \cite{ji2024wavtokenizer} (75 tokens/s, single 4096-code codebook, 24\,kHz) as the tokenizer for a sample of the bioacoustic data in BirdCLEF 2026.
Perch v2 served as the round-trip evaluator.
We evaluated this experiment on round-trip quality against Perch embeddings and logits, alongside semantic retrieval tasks.
The first task is a round-trip reconstruction task where we take a clip of audio, pass it through the codec encoder-decoder, and compare the cosine distance of the Perch embeddings under lossless and lossy conditions.
The second task uses the same round-tripped audio, but this time does a retrieval task over the Perch embeddings.
The third task uses the WavTokenizer pooled embeddings directly for retrieval on focal recordings.
We choose a set of 10 species to analyze the performance.

\input{sections/floats/51_tab_wavtokenizer-gates}

WavTokenizer codec retrieval failed on every measured account (Table~\ref{tab:wavtokenizer-gates}).
Round-tripping audio through the codec collapses Perch-embedding cosine similarity (0.313 against the 0.90 gate) and top-1 species accuracy (9\%).
The resulting encodings are also not useful for semantic search in this domain.
WavTokenizer was trained on human speech, and breaking round-trip performance down by species shows that lower-frequency, speech-like vocalizations survive the codec better than high-frequency bird calls (Table~\ref{tab:wavtokenizer-roundtrip}).

\input{sections/floats/51_tab_wavtokenizer-roundtrip}

\subsection{Semantic Representations}

We measure how well foundational audio embeddings transfer by ranking encoders on the BirdCLEF target task.
We evaluate encoders built largely on vision-transformer architectures, which have not typically appeared as competitive solutions.
We screened these encoders on ESC-50 \cite{piczak2015esc50} via 5-fold linear probing and separately probed a static-embedding deployment path for cheaper CPU inference; both are collected in Appendix~\ref{sec:appendix-encoders}.
We finetune with LoRA adapters \cite{hu2022lora} as a parameter-efficient alternative to full updates, since many of these models are large and require significant GPU memory.
ESC-50 is a proxy task, environmental sound classification, and does not rank these encoders on the target soundscape task; however it is a useful source reproducibility check.

AST \cite{gong2021ast} is a strong spectrogram-transformer baseline at 95.5\% on ESC-50, and one of the first to experiment with vision transformers in the audio domain.
BEATs \cite{chen2023beats} is the strongest encoder we tested at 95.8\%, and is trained using self-supervision and masking.
EAT \cite{chen2024eat} is an efficient audio transformer that extends the BEATs masked-prediction recipe with more aggressive masking and benefits from LoRA finetuning.
SSAMBA \cite{shams2024ssamba} is a variation of AST that uses Mamba layers \cite{gu2023mamba} to approximate attention and can in theory scale to much larger token regimes than AST can.

To compare these models, we fit a ridge \cite{hoerl1970ridge} head on each encoder's frozen embeddings using the focal training recordings.
We evaluate using a leave-one-site-out (LOSO) split over the 739 labeled windows and report pooled out-of-fold macro AUROC.
This validation metric is different from those in the CNN baseline, but their Kaggle leaderboard measures are provided for consistency in the final test set.
We add the bioacoustic specialists Perch v2 \cite{hamer2025perch2} and BirdNET \cite{kahl2021birdnet} as anchors.

\subsection{Leaderboard Results}

\input{sections/floats/52_tab_birdclef2026-loso}

We report LOSO training scores and post-competition leaderboard results in Table~\ref{tab:birdclef2026-loso}.
In local scoring, Perch as a bioacoustic specialist barely beats the general encoder BEATs on frozen features, while AST, EAT, and SSAMBA cluster at the bottom.
However, on the leaderboard, Perch far outranks the other models, with the other specialist BirdNET coming in second.
BirdNET overtakes BEATs despite ranking below it offline, so our offline ranking is not necessarily a good predictor of leaderboard standing.
Perch is also the most robust to the public-to-private shift, while the SSL-only SSAMBA degrades the most.

%% file: sections/floats/51_tab_wavtokenizer-gates.tex
\begin{table}[h]
  \centering
  \caption{
  WavTokenizer codec retrieval against its predefined success gates.
  Every gate is missed by a wide margin.
  The soundscape phase carried no hard gate; its recall, taken at the most permissive threshold, is indistinguishable from noise.
  }
  \label{tab:wavtokenizer-gates}
  \begin{tabular}{llccc}
    \toprule
    Phase              & Metric                 & Target     & Result & Met? \\
    \midrule
    Round-trip         & Perch embedding cosine & >0.90    & 0.313  & $\times$ \\
    Round-trip         & Precision@1        & >0.80    & 0.09  & $\times$ \\
    Focal retrieval    & Precision@10                   & >0.80    & 0.195  & $\times$ \\
    \bottomrule
  \end{tabular}
\end{table}

%% file: sections/floats/51_tab_wavtokenizer-roundtrip.tex
\begin{table}[h]
  \centering
  \caption{
  Per-species round-trip cosine similarity.
  Speech-like vocalizations (low-frequency tonal calls, dog barks) survive the codec better than the high frequency bird calls.
  }
  \label{tab:wavtokenizer-roundtrip}
  \begin{tabular}{llrr}
    \toprule
    Species                    & Class    & Cos sim & Top-1 \\
    \midrule
    Lesser Snouted Tree Frog   & Amphibia & 0.539   & 25\%  \\
    Domestic Dog               & Mammalia & 0.423   & 10\%  \\
    Pale-legged Weeping Frog   & Amphibia & 0.374   & 5\%   \\
    Purplish Jay               & Aves     & 0.357   & 5\%   \\
    Common Potoo               & Aves     & 0.340   & 30\%  \\
    Bare-faced Curassow        & Aves     & 0.337   & 15\%  \\
    Rufous Hornero             & Aves     & 0.214   & 0\%   \\
    Great Kiskadee             & Aves     & 0.201   & 0\%   \\
    Chestnut-vented Conebill   & Aves     & 0.200   & 0\%   \\
    White-faced Whistling Duck & Aves     & 0.145   & 0\%   \\
    \bottomrule
  \end{tabular}
\end{table}

%% file: sections/floats/52_tab_birdclef2026-loso.tex
\begin{table}[h]
  \centering
  \caption{
    Frozen LOSO ranking vs.\ deployed Kaggle leaderboard for six encoders on BirdCLEF 2026.
    AUROC is the leaderboard-aligned score from the 9-fold LOSO ridge probe; the public/private LB deploys the same ridge head.
    CPU min/1k is the projected minutes per 1000 60\,s clips at 4 vCPU.
  }
  \label{tab:birdclef2026-loso}
  \begin{tabular}{lrrrrr}
    \toprule
            &      & LOSO  & \multicolumn{2}{c}{Kaggle LB} & CPU            \\
    \cmidrule(lr){4-5}
    Encoder & dim  & AUROC & pub            & priv         & min/1k         \\
    \midrule
    Perch   & 1536 & \textbf{0.620} & \textbf{0.866} & \textbf{0.862} & \textbf{11.5} \\
    BEATs   & 768  & 0.612          & 0.777          & 0.760          & 29.6          \\
    BirdNET & 1024 & 0.588          & 0.814          & 0.798          & 34.5          \\
    AST     & 768  & 0.576          & 0.778          & 0.768          & 38.2          \\
    EAT     & 768  & 0.569          & 0.780          & 0.766          & 25.5          \\
    SSAMBA  & 384  & 0.563          & 0.731          & 0.689          & 72.5          \\
    \bottomrule
  \end{tabular}
\end{table}

%% file: sections/60_discussion.tex
\section{Discussion}

\subsection{The Geometry of Semantic Tokens}

\input{sections/floats/52_fig_separability}

To analyze a feature encoder, we visualize how classes cluster in space.
Our results are consistent with \cite{ghani2023global}, where birdsong embeddings out-compete AudioMAE and show better separability in a t-SNE scatterplot of their embeddings.
In Figure~\ref{fig:encoder-separability}, we apply PaCMAP to recording centroids across a small hand-selected set of animals.
The specialists are nearly linearly separable, while transformer-based models conflate several bird species.
SSAMBA pools everything together into a blob.
The leaderboard ordering is partly recoverable from $k$-nearest-neighbor accuracy over recording centroids used in the embedding plot ($k=10$, macro-averaged over the 206 focal species).
This kNN ranking recovers the top and bottom of the public leaderboard (Perch 0.71, BirdNET 0.58, AST 0.29, EAT 0.27, BEATs 0.23, SSAMBA 0.07).

\subsection{Toward Sequence Modeling of Arbitrary-Length Audio}

Our experiments and successful solutions on the leaderboard demonstrate that the strongest practical systems are the ones that take advantage of foundational bioacoustic backbones and exploit the newly added supervision signals.
These sample-efficient methods use strong foundational encoders and adjust for the domain shift, but all rely on some labeling, weak or otherwise.
Between Perch and EAT, there is nearly a 9-point gap in public-leaderboard macro AUROC, reflecting Perch's stronger discriminative power across taxa and species, and Perch is also over 2x faster at inference.
The differences are no surprise, given that transformer-based models lack the inherent spatial and temporal biases of convolutional neural networks and are forced into token embeddings that were designed to make discrete data continuous.
Additionally, the computational complexity of attention forces engineering decisions that result in a Pareto frontier of potential solutions that scale with the available data and compute.

Our experimental approach this year failed to exploit the unlabeled soundscapes, with which the labeled soundscapes are distributionally aligned.
As in many previous competition years, there is a domain shift between the focal recordings, which contain clear indications of a particular animal's vocalization, and a soundscape, where there can be many different vocalizations at the same time at different distances.
Transformer-based models can exploit unlabeled data through self-supervised objectives like masked reconstruction.
The Bird-MAE authors show that, when trained correctly, the larger parameterization of transformer-based models trained on a specific domain can outperform models like Perch by a fair margin.
However, a limitation of their training methods is that these models were given a limited context of audio up to 10 seconds, which aligns with the BirdSet (or AudioSet) classification tasks.

\input{sections/floats/60_fig_tokenization}

One potential path forward would be to build foundational models that can ingest arbitrary length audio instead of fixed 5-10 second windows.
Figure~\ref{fig:tokenization} contrasts two tokenization regimes: the default ViT $16\times16$ square patches fix the model into a two-dimensional grid, whereas full-band $128\times1$ columns yield a one-dimensional token stream that can be ingested into a sequence model.
The transformer-based audio models like AST describe some limited experiments with tokenization strategies that favor taller tokens spanning more of the frequency range which achieve comparable or better classification performance, but many models found in shared repositories distribute square patches due to the computational savings from starting off a backbone initialized on something like ImageNet.
This representation is well suited to linear-time backbones like Mamba, or other models that scale linearly with the number of tokens, in their ability to contextualize a whole soundscape instead of just the immediate fragment under consideration.
We experimented with continued self-supervised learning of a pre-trained SSAMBA \cite{shams2024ssamba}.
At the scale we could train, however, the result was inconclusive: our 55k-clip self-supervised run is an order of magnitude smaller than SSL runs in the literature, and none of our models reached the supervised BirdSet \cite{rauch2025birdset} baselines (Appendix~\ref{sec:appendix-ssamba}).

Despite there being 177 hours of soundscape audio, these are distributed across a much smaller number of sites.
The time and place then becomes a useful deductive tool that could be implicitly encoded into a backbone.
Another natural direction is to encourage few-shot behavior by creating question and answer pairs joined by separator tokens, and to propagate back signal through a head that learns to attend to similar patterns.
This might exhibit itself as a prototypical retrieval system where classes with only a few examples can be used to modulate how the model perceives tokens in context.
Additionally, it might be viable to build a multi-modal model where language describing the taxonomy and the audible properties of the soundscape could be trained with a CLIP loss to aid in active learning loops.
Despite the leaderboard's affinity for small, fast models, larger and more expressive models might still exploit low-resource class data well enough to distill into a smaller deployable model.

\section{Conclusion}

This year we pursued a supervised baseline built to the community's best practices, and a study of whether token-based representations could compete with it on the same leaderboard.
The baseline carried the result, with an equal-weight ensemble of a frozen Perch v2 backbone and a trained HGNetV2-B0 sound-event-detection network reaching 0.929 at rank 1962 on private leaderboard within the 90-minute CPU budget.
Adding in a prototypical head for insects brough this up to 0.936 at rank 1894.
The representational thread was less conclusive.
The codec route failed on bioacoustic retrieval due to representational constraints, the semantic embeddings transferred only modestly due to domain gap, and the linear-time state-space alternative remained inconclusive at the scale we could train.
However, we look forward to foundational bioacoustic systems that exploit unlabeled data and make use of low-resource classes, in a domain where expertise is limited and conservation needs are urgent.
We are curious to see how artificial intelligence develops in service of conservation, ideally in ways that help us quantify and act on positive change.


%% file: sections/floats/52_fig_separability.tex
\begin{figure}[h]
  \centering
  \includegraphics[width=\textwidth]{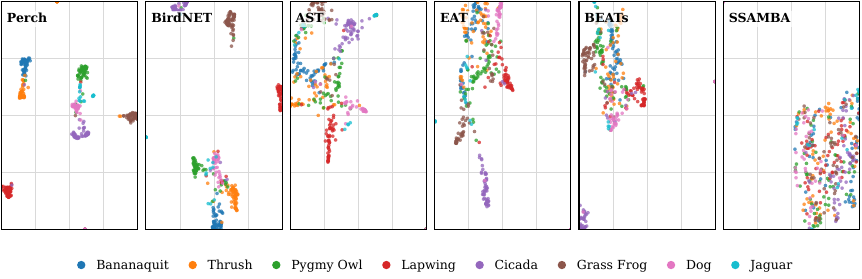}
  \caption{
    Embedding-space species separability across the six deployed encoders.
    Two-dimensional PaCMAP projections of frozen embeddings for eight BirdCLEF 2026 species (one point per focal recording, mean-pooled).
  }
  \label{fig:encoder-separability}
\end{figure}

%% file: sections/floats/60_fig_tokenization.tex
\definecolor{gridGray}{HTML}{9AA0A6}
\definecolor{streamBlue}{HTML}{7FB2D8}
\definecolor{inkBlue}{HTML}{17608F}
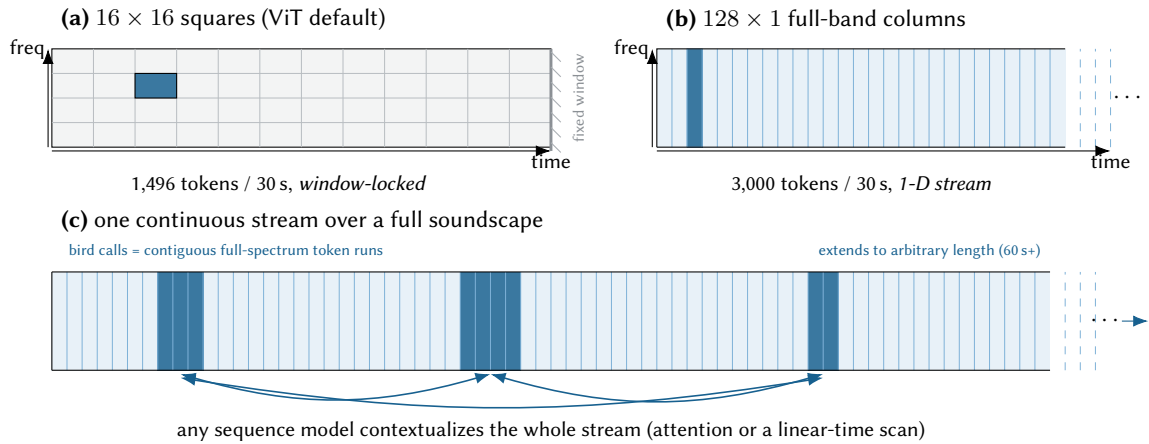
\begin{figure}[t]
  \centering
  \begin{tikzpicture}[
      font=\footnotesize,
      >={Latex[length=2mm]},
      spec/.style={draw=black, thin},
      gridline/.style={draw=gridGray!70, very thin},
      colline/.style={draw=streamBlue, very thin},
      tokhi/.style={fill=inkBlue!85},
      wall/.style={draw=gridGray!90!black, line width=1.1pt},
      ttl/.style={anchor=south west},
      lbl/.style={align=center, font=\scriptsize},
      ctx/.style={<->, draw=inkBlue, semithick},
    ]
    \node[ttl] at (0,1.42) {\textbf{(a)} $16\times16$ squares (ViT default)};
    \draw[spec, fill=gridGray!12] (0,0) rectangle (6.6,1.3);
    \foreach \i in {1,...,11}{\draw[gridline] (\i*0.55,0) -- (\i*0.55,1.3);}
    \foreach \j in {1,2,3}{\draw[gridline] (0,\j*0.325) -- (6.6,\j*0.325);}
    \fill[tokhi] (1.1,0.65) rectangle (1.65,0.975);
    \draw[spec]  (1.1,0.65) rectangle (1.65,0.975);
    \draw[->] (0,-0.05) -- (6.6,-0.05) node[below,font=\scriptsize,inner sep=1pt]{time};
    \draw[->] (-0.05,0) -- (-0.05,1.3) node[left,font=\scriptsize,inner sep=1pt]{freq};
    \draw[wall] (6.6,-0.05) -- (6.6,1.3);
    \foreach \y in {0.06,0.26,...,1.28}{\draw[gridGray!85, thin] (6.6,\y) -- (6.73,\y-0.13);}
    \node[font=\tiny, text=gridGray!90!black, rotate=90] at (6.96,0.65) {fixed window};
    \node[lbl, anchor=north] at (3.0,-0.25)
      {1{,}496 tokens / 30\,s, \emph{window-locked}};
    \node[ttl] at (8.0,1.42) {\textbf{(b)} $128\times1$ full-band columns};
    \fill[streamBlue!18] (8.0,0) rectangle (13.4,1.3);
    \foreach \i in {1,...,26}{\draw[colline] (8.0+\i*0.2,0) -- (8.0+\i*0.2,1.3);}
    \fill[tokhi] (8.4,0) rectangle (8.6,1.3);
    \draw[colline] (8.4,0) rectangle (8.6,1.3);
    \draw[spec] (8.0,0) -- (8.0,1.3) -- (13.4,1.3);
    \draw[spec] (8.0,0) -- (13.4,0);
    \foreach \i in {0,1,2}{\draw[colline, dashed] (13.6+\i*0.2,0) -- (13.6+\i*0.2,1.3);}
    \node at (14.25,0.65) {$\cdots$};
    \draw[->] (8.0,-0.05) -- (14.0,-0.05) node[below,font=\scriptsize,inner sep=1pt]{time};
    \draw[->] (7.94,0) -- (7.94,1.3) node[left,font=\scriptsize,inner sep=1pt]{freq};
    \node[lbl, anchor=north] at (10.7,-0.25)
      {3{,}000 tokens / 30\,s, \emph{1-D stream}};
    \node[ttl] at (0,-1.25) {\textbf{(c)} one continuous stream over a full soundscape};
    \node[lbl, font=\tiny, text=inkBlue, anchor=south west] at (0.1,-1.6)
      {bird calls = contiguous full-spectrum token runs};
    \node[lbl, font=\tiny, text=inkBlue, anchor=south east] at (13.2,-1.6)
      {extends to arbitrary length (60\,s+)};
    \fill[streamBlue!18] (0,-2.95) rectangle (13.2,-1.65);
    \foreach \x in {1.4,1.6,1.8}{\fill[tokhi] (\x,-2.95) rectangle (\x+0.2,-1.65);}
    \foreach \x in {5.4,5.6,5.8,6.0}{\fill[tokhi] (\x,-2.95) rectangle (\x+0.2,-1.65);}
    \foreach \x in {10.0,10.2}{\fill[tokhi] (\x,-2.95) rectangle (\x+0.2,-1.65);}
    \foreach \i in {1,...,65}{\draw[colline] (\i*0.2,-2.95) -- (\i*0.2,-1.65);}
    \draw[spec] (0,-1.65) -- (13.2,-1.65);
    \draw[spec] (0,-2.95) -- (13.2,-2.95);
    \draw[spec] (0,-2.95) -- (0,-1.65);
    \foreach \i in {0,1,2}{\draw[colline, dashed] (13.4+\i*0.2,-2.95) -- (13.4+\i*0.2,-1.65);}
    \node at (13.95,-2.3) {$\cdots$};
    \draw[->,inkBlue] (14.15,-2.3) -- (14.5,-2.3);
    \draw[ctx] (1.7,-3.0) to[bend right=16] (5.8,-3.0);
    \draw[ctx] (5.8,-3.0) to[bend right=16] (10.2,-3.0);
    \draw[ctx] (1.7,-3.05) to[bend right=9] (10.2,-3.05);
    \node[lbl, anchor=north] at (6.6,-3.5)
      {any sequence model contextualizes the whole stream (attention or a linear-time scan)};
  \end{tikzpicture}
  \caption{
    Patch geometry on ViT-styled tokenization of spectrograms.
    \textbf{(a)} The default ViT grid ties tokens to a fixed 2-D layout where audio is locked into dimensions of the training data.
    \textbf{(b)} Full-band $128\times1$ columns collapse the entire spectrum into each token ($128$ mel bins $\times$ 10\,ms, one per frame) yielding a 1-D stream.
    \textbf{(c)} Because the stream is one-dimensional, a sequence model can contextualize an entire soundscape the way a language model reads a long prompt.}
  \label{fig:tokenization}
\end{figure}

%% file: sections/70_closing.tex
\section*{Acknowledgments}

We thank the Data Science at Georgia Tech Applied Research Competitions group (DS@GT ARC) for their support.
This research was supported in part through research cyberinfrastructure resources and services provided by the Partnership for an Advanced Computing Environment (PACE) at the Georgia Institute of Technology, Atlanta, Georgia, USA \cite{PACE}.

\section*{Declaration on Generative AI}

AI-assisted tools (e.g., Claude Code, Codex) were used during the development of this work to aid in programming, debugging, grammar and spelling checks, literature review, and experimentation. 
All experiments were designed and configured by the authors, all results were manually verified, and all scientific interpretations and conclusions reflect the authors’ own judgment. 
The authors take full responsibility for the accuracy, integrity, and originality of the work presented in this paper.

%% file: sections/90_appendix_baseline.tex
\section{Baseline Ablation Record}
\label{sec:appendix-baseline}

This appendix records the experiments behind the supervised baseline (Section~\ref{sec:baseline}): an overview of every main experiment, the parity of three SED recipes, and the curated negative results that justify the deliberately simple deployed model.

Table~\ref{tab:experiment-overview} summarizes the main experiments and their leaderboard results.

\input{sections/floats/90_tab_experiment-overview}

The deployed HGNetV2-B0 SED is the simplest of three recipes that reach the same public LB (Table~\ref{tab:baseline-parity}).
An ImageNet EfficientNet-B0 SED distilled from Perch v2 embeddings recovers the identical single-fold score (a $+0.050$ gain over its no-distillation control), and an EfficientNetV2-S pretrained from scratch on broad Neotropical Xeno-Canto audio also matches it.
Since distillation requires a teacher pipeline and custom pretraining adds a large-scale training stage, neither buys accuracy over the plain ImageNet recipe.

\input{sections/floats/90_tab_baseline-parity}

The discipline behind the ladder in Section~\ref{sec:baseline} is that offline validation AUROC anti-correlated with the public leaderboard across 20$+$ submissions, so every change was treated as a single-variable experiment arbitrated by a Kaggle submission.
Mst candidate improvements regressed (Table~\ref{tab:baseline-negatives}).
The levers that worked were few and frontend- or architecture-level.

\input{sections/floats/90_tab_baseline-negatives}

%% file: sections/floats/90_tab_experiment-overview.tex
\begin{table}[h]
  \centering
  \footnotesize
  \caption{
    Overview of the main experiments run for the supervised system, grouped by track.
    Each row is the best public-LB (macro-AUROC) result for that experiment.
    Roles: \emph{deployed} (final submission), \emph{ensemble} (member or its lineage),
    \emph{explored} (matched but not adopted), \emph{superseded} (improved on later),
    \emph{rejected} (regressed or no gain).
  }
  \label{tab:experiment-overview}
  \begin{tabular}{@{}lrrl@{}}
    \toprule
    Approach / model                              & Exp & pub LB         & Role        \\
    \midrule
    \multicolumn{4}{l}{\emph{Clip-level fine-tuning (LoRA encoders)}} \\
    \quad ProtoCLR (focal)                        & 011 & 0.736          & superseded  \\
    \quad ProtoCLR ($+$ labeled soundscape)       & 012 & 0.767          & superseded  \\
    \quad ProtoCLR ($+$ pseudo-labels)            & 015 & 0.762          & rejected    \\
    \midrule
    \multicolumn{4}{l}{\emph{Sound-event detection (SED)}} \\
    \quad EfficientNet-B0 SED                     & 016 & 0.827          & superseded  \\
    \quad $+$ high-res mel $+$ AmplitudeToDB      & 017 & 0.859          & superseded  \\
    \quad HGNetV2-B0 SED (4-fold)                 & 019 & 0.913          & ensemble    \\
    \quad $+$ site/hour priors                    & 021 & 0.917          & ensemble    \\
    \midrule
    \multicolumn{4}{l}{\emph{Frozen Perch~v2 probes}} \\
    \quad $+$ MLP probes (OOF)                    & 018 & 0.918          & ensemble    \\
    \quad $+$ ProtoSSM temporal head (TF)         & 020 & 0.925          & explored    \\
    \midrule
    \multicolumn{4}{l}{\emph{Pretraining \& distillation}} \\
    \quad EfficientNet-B0 SED $+$ Perch KD (5-fold) & 032 & 0.915        & explored    \\
    \quad EfficientNetV2-S broad-XC pretrain (4-fold) & 038 & 0.916      & explored    \\
    \midrule
    \multicolumn{4}{l}{\emph{Other directions explored}} \\
    \quad PCEN normalization                      & 022 & 0.880          & rejected    \\
    \quad Iterative pseudo-labeling               & 025 & 0.897          & rejected    \\
    \quad Alternate losses / inference            & 027 & 0.906          & rejected    \\
    \quad 2025$\to$2026 warm-start                & 029 & 0.885          & rejected    \\
    \quad Sydorskyi mixup                         & 030 & 0.901          & rejected    \\
    \quad Own-teacher pseudo-labels               & 033 & 0.899          & rejected    \\
    \quad Noisy-student soft pseudo-labels        & 037 & 0.832          & rejected    \\
    \midrule
    \multicolumn{4}{l}{\emph{Final ensemble}} \\
    \quad Perch~v2 probe $+$ HGNetV2-B0 SED       & 023 & \textbf{0.929} & deployed    \\
    \bottomrule
  \end{tabular}
\end{table}

%% file: sections/floats/90_tab_baseline-parity.tex
\begin{table}[h]
  \centering
  \caption{
    Three independent SED recipes converge to the same public LB.
    Knowledge distillation and broad-Xeno-Canto pretraining match, but do not beat, the plain ImageNet HGNetV2-B0 recipe we deploy.
  }
  \label{tab:baseline-parity}
  \begin{tabular}{llrr}
    \toprule
    Recipe                          & Pretraining         & single & 4-fold        \\
    \midrule
    HGNetV2-B0 SED (deployed)       & ImageNet            & 0.906  & 0.917         \\
    EfficientNet-B0 SED $+$ Perch KD & ImageNet            & 0.906  & 0.915         \\
    EfficientNetV2-S SED            & broad Xeno-Canto    & 0.907  & 0.916         \\
    \bottomrule
  \end{tabular}
\end{table}

%% file: sections/floats/90_tab_baseline-negatives.tex
\begin{table}[h]
  \centering
  \caption{
    Curated negative results for the SED member.
    Across 20$+$ single-variable leaderboard submissions, offline validation AUROC anti-correlated with the public LB, and nearly every training-recipe or extra-data change regressed.
    The levers that worked were few and frontend- or architecture-level (Table~\ref{tab:sed-ladder}).
  }
  \label{tab:baseline-negatives}
  \begin{tabular}{lrl}
    \toprule
    Change                            & $\Delta$ LB          & Root cause                              \\
    \midrule
    Pseudo-labeling (6 variants)      & $-0.005$ to $-0.074$ & diffuse soft labels poison training     \\
    Cross-entropy loss                & $-0.41$              & logits collapse without per-class margin \\
    Waveform mixup                    & $-0.019$ to $-0.036$ & amplitude shift, train/infer mismatch   \\
    Learnable PCEN                    & $-0.026$             & inferior to AmplitudeToDB               \\
    Attention pooling head            & $-0.013$ to $-0.035$ & train/infer window mismatch             \\
    Multi-context (10--20\,s)         & $-0.010$ to $-0.025$ & window-length mismatch                  \\
    2025$\to$2026 warm-start          & $-0.021$             & under-trained random-init head rows     \\
    Training-recipe sweep             & 0-for-14             & v5j at a local optimum                  \\
    \bottomrule
  \end{tabular}
\end{table}

%% file: sections/90_appendix_bands.tex
\section{Taxon Frequency Overlap}
\label{sec:appendix-bands}

The spectrogram frontend (Section~\ref{sec:baseline}) is motivated by the taxa not being separable by frequency band.
We measure band occupancy over 80 random recordings per taxon (Figure~\ref{fig:band-occupancy}).
Frogs, birds, and insects all concentrate energy in 1--4\,kHz, with peak-frequency medians of 1.8, 2.9, and 1.8\,kHz.
A hand-crafted frequency-band feature cannot separate the taxa (Figure~\ref{fig:taxa-spectrograms}).

\input{sections/floats/90_fig_band-occupancy}

%% file: sections/floats/90_fig_band-occupancy.tex
\begin{figure}[h]
  \centering
  \includegraphics[width=\textwidth]{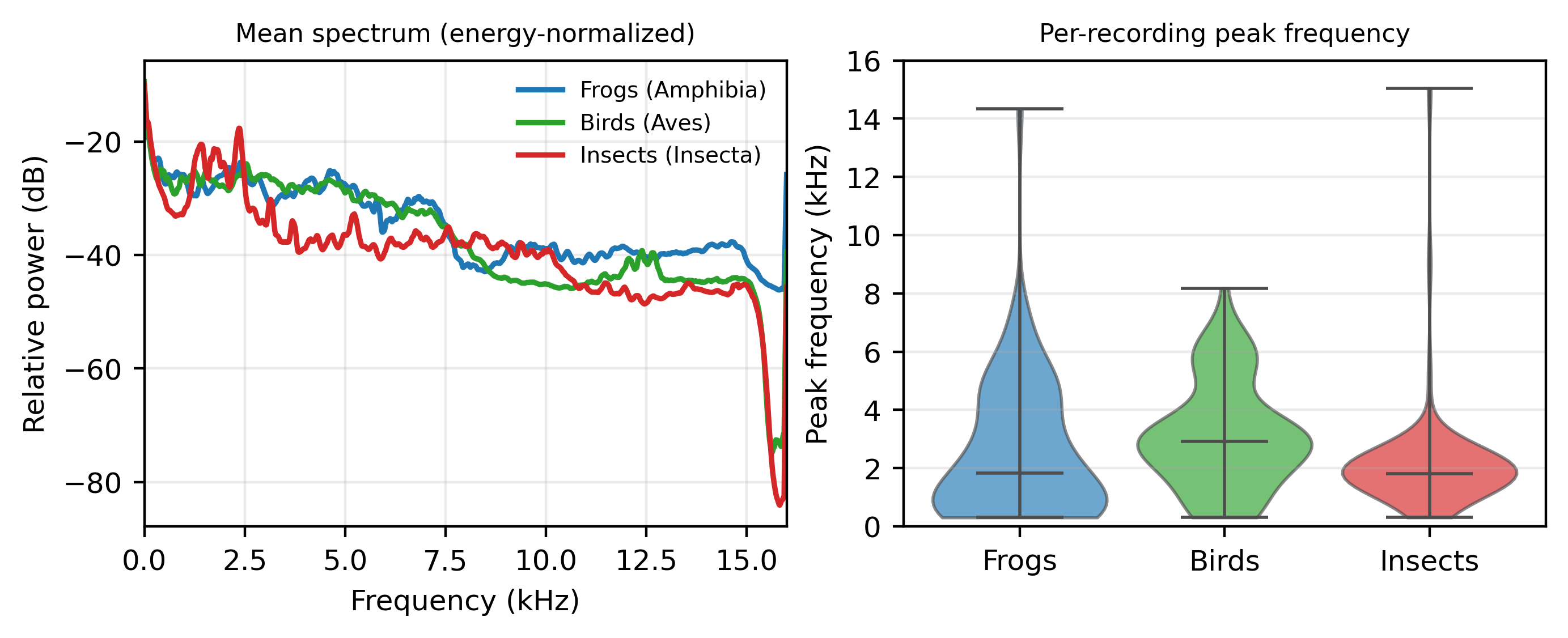}
  \caption{
    Spectral band occupancy by taxon, over 80 random recordings each.
    Left: mean energy-normalized power spectrum.
    Right: per-recording peak frequency, with the median marked.
    All three taxa concentrate energy in 1--4\,kHz (peak-frequency medians: frogs 1.8, birds 2.9, insects 1.8\,kHz), so frequency band alone does not separate them.
  }
  \label{fig:band-occupancy}
\end{figure}

%% file: sections/90_appendix_perch.tex
\section{Frozen Perch Transfer Learning}
\label{sec:appendix-perch}

This appendix documents exploratory probes on frozen Perch v2 \cite{hamer2025perch2} embeddings that model the focal-to-soundscape domain shift directly.
They add considerable modeling machinery for a result that did not surpass the supervised baseline. 

All heads operate on Perch v2's embeddings and are evaluated under a 9-fold leave-one-site-out (LOSO) split on the labeled soundscape windows, reporting macro AUROC and cmAP.
We compare three heads (Table~\ref{tab:perch-heads}):
\begin{itemize}
  \item \textbf{Ridge + empirical-Bayes prior.} Ridge regression \cite{hoerl1970ridge} with a per-class L2 penalty and Bayesian-style pooling that shares information across the species and site dimensions.
  \item \textbf{Star-schema prototype retrieval.} A recommendation-system framing of the problem with a species fact table on an item axis, a site dimension table on a context axis, and feedback observed on the (window, species, site) join, scored by cosine similarity against multiple prototypes.
  \item \textbf{Hopfield retrieval.} A modern Hopfield head \cite{ramsauer2021hopfield} as a learned prototype-retrieval layer over a representation shared across focal and soundscape recordings.
\end{itemize}

\input{sections/floats/90_tab_perch-heads}

\input{sections/floats/90_tab_perch-starschema}

\input{sections/floats/90_tab_perch-hopfield-subsets}

We find the domain shift is real and has to be accounted for explicitly.
Prototype retrieval in Perch's embedding space works well on species with focal recordings (Table~\ref{tab:perch-starschema}) but collapses to chance for the 28 species with none.
A learned Hopfield projection recovers part of that gap ($0.258 \to 0.661$, Table~\ref{tab:perch-hopfield-subsets}), while ridge regression sidesteps it by regularizing toward the mean.
Insect prototypes are unreachable, as 25 of 28 do not have focal data or taxonomic siblings.
Characterizing this shift more carefully is ongoing work.
Recent benchmarks like BirdSet \cite{rauch2025birdset} formalize the focal-to-soundscape transfer problem, and the DCASE bioacoustic task documents a similar gap \cite{ghani2024domaingap}.

%% file: sections/floats/90_tab_perch-heads.tex
\begin{table}[h]
  \centering
  \caption{
    Perch heads on the labeled soundscape subset.
    These AUROC values use an empirical-Bayes prior estimator and are not directly comparable to the leaderboard-aligned pooled-OOF figure for Perch in Table~\ref{tab:birdclef2026-loso} (0.620).
  }
  \label{tab:perch-heads}
  \begin{tabular}{lr}
    \toprule
    Method                                    & AUROC                                  \\
    \midrule
    \quad Ridge + empirical-Bayes prior       & 0.749                                  \\
    \quad Star schema cosine, multi-prototype & 0.696                                  \\
    \quad Hopfield, learned projection        & 0.761                       \\
    \bottomrule
  \end{tabular}
\end{table}

%% file: sections/floats/90_tab_perch-starschema.tex
\begin{table}[h]
  \centering
  \caption{
    Star schema decomposition by focal data availability.
    Prototype retrieval in Perch's embedding space works on species with focal recordings but collapses to random for species without.
    Ridge regression regularizes predictions toward the mean and does not see the same gap.
  }
  \label{tab:perch-starschema}
  \begin{tabular}{lrrr}
    \toprule
    Model                        & w/ focal (42 sp) & w/o focal (28 sp) & all (70 sp)    \\
    \midrule
    Star schema cosine           & \textbf{0.816}     & 0.258 (random)        & 0.593          \\
    Ridge (focal emb + EB prior) & 0.858              & 0.500 (random)        & \textbf{0.749} \\
    \bottomrule
  \end{tabular}
\end{table}

%% file: sections/floats/90_tab_perch-hopfield-subsets.tex
\begin{table}[h]
  \centering
  \caption{
  Hopfield factored retrieval per-subset breakdown.
  The learned projection lifts the focal-less subset from 0.258 (Table~\ref{tab:perch-starschema}) to 0.661.
  Insect morphotypes remain unreachable: 25 of 28 are unresolved sound-types with no focal data and no taxonomic siblings.
  }
  \label{tab:perch-hopfield-subsets}
  \begin{tabular}{lrr}
    \toprule
    Subset                       & Species & AUROC            \\
    \midrule
    Birds (data-rich)            & 162     & \textbf{0.868}   \\
    Amphibians (medium-resource) & 35      & 0.653            \\
    Insect morphotypes           & 28      & --- $^{\dagger}$ \\
    \midrule
    With focal data              & 206     & 0.769            \\
    Without focal data           & 28      & 0.661            \\
    \bottomrule
  \end{tabular}

  \vspace{0.5em}
  \footnotesize{$^{\dagger}$No positive labels in the held-out fold; AUROC undefined.}
\end{table}

%% file: sections/90_appendix_encoders.tex
\section{Encoder Screening and Static-Embedding Deployment}
\label{sec:appendix-encoders}

This appendix collects the general-domain encoder screening and a static-embedding deployment.
We evaluate each encoder through a proxy task that exploits the learned linearity and separability of the representation.
We use 5-fold linear probing on ESC-50 \cite{piczak2015esc50}, a small environmental sound classification dataset that can be evaluated quickly and has concrete classes usable in a supervised training pipeline.
Table~\ref{tab:encoder-benchmark} reports 5-fold accuracy across encoders and training strategies (linear probe, LoRA, full finetune).

\input{sections/floats/90_tab_encoder-benchmark}

\input{sections/floats/90_tab_beats-deployment}

Beyond fine-tuning, we tested whether static embeddings could reduce inference cost.
We use the Model2Vec methodology from language models which takes a static codebook of tokens, applies smooth inverse frequency (SIF) to reweight principal components of the tokens, and adds a linear probe at the head.
Passing the entire vocabulary through the sequence model alone captures enough contextual knowledge that cross-attention can be dropped at inference, often an order-of-magnitude speedup on CPU.
We also experimented with distilling a token model, such that we could construct the original tokens using the 500k AudioSet subset \cite{gemmeke2017audioset} to match the per-position contextual embeddings.
We used BEATs as the teacher and ConvNeXt1D-Small (2.5M parameters) as the student, but poor GPU utilization led us to drop this experiment.

Static embeddings gave an order-of-magnitude speedup at a substantial accuracy cost (Table~\ref{tab:beats-deployment}).
Several recontextualization methods on the target task failed to recover the lower bound of full-model performance.

%% file: sections/floats/90_tab_encoder-benchmark.tex
\begin{table}[h]
  \centering
  \caption{ESC-50 5-fold CV accuracy across encoders and training strategies.}
  \label{tab:encoder-benchmark}
  \begin{tabular}{llc}
    \toprule
    Encoder         & Strategy                       & Accuracy (\%)    \\
    \midrule
    BEATs           & Linear probe                   & \textbf{95.8}    \\
    AST             & Linear probe                   & 95.5             \\
    EAT             & Linear probe                   & 77.9             \\
    EAT             & LoRA $r{=}8$                   & 94.6             \\
    SSAMBA-HF-small & Linear probe                   & 61.6             \\
    SSAMBA-HF-small & LoRA $r{=}8$ (matched recipe)  & 87.0$^{\dagger}$ \\
    SSAMBA-HF-small & Full finetune (matched recipe) & 87.0$^{\dagger}$ \\
    wav2vec2        & Linear probe                   & 48.2             \\
    wav2vec2        & LoRA $r{=}8$                   & 3.0              \\
    \bottomrule
  \end{tabular}

  \vspace{0.5em}
  \footnotesize{$^{\dagger}$Fold 1 only; full 5-fold CV did not complete.}
\end{table}

%% file: sections/floats/90_tab_beats-deployment.tex
\begin{table}[h]
  \centering
  \caption{
    ESC-50 5-fold CV accuracy and CPU inference speedup for BEATs deployment heads.
    Replacing the BEATs forward pass with a static token-codebook lookup gives a $\sim$17$\times$ CPU speedup but loses 48 points of accuracy.
    Feature-space transforms recover a few points; learned poolers over the static lookup recover some of the gap, but does not close it.
    Speedup is wall-clock CPU inference vs full BEATs (1503\,s for 2000 ESC-50 clips, 4-thread Intel Xeon Gold 6226).
    For learned pooler heads, we report partial results only 2 out of the 5 folds.
    }
  \label{tab:beats-deployment}
  \begin{tabular}{llrrr}
    \toprule
    Configuration            & Head                               & Params & Acc (\%)         & Speedup \\
    \midrule
    Full BEATs (teacher)     & Mean pool + Linear                 & 90M    & \textbf{84.0}    &
    1.0$\times$                                                                                         \\
    \midrule
    \multicolumn{5}{l}{\textit{Static codebook lookup, frozen head}}                                    \\
    \quad 768d               & Mean pool                          & 0      & 35.9             &
    17$\times$                                                                                          \\
    \quad 768d + SIF         & SIF-weighted mean                  & 0      & 35.7             &
    16$\times$                                                                                          \\
    \quad 256d (PCA + SIF)   & SIF-weighted mean                  & 0      & 43.9             &
    17$\times$                                                                                          \\
    \midrule
    \multicolumn{5}{l}{\textit{Static codebook lookup, learned pooler head}}                            \\
    \quad AttentionPooler    & Attention + mean                   & 1.5M   & 38.9             &
    ---                                                                                    \\
    \quad CNNAttentionPooler & CNN + attention                    & 2.0M   & 43.4             &
    ---                                                                                    \\
    \quad LightCNN1DPooler   & 768$\to$256$\to$768 bottleneck CNN & 1.0M   & 58.1             &
    ---                                                                                    \\
    \quad CNN1DPooler        & 768$\to$768 CNN                    & 5.9M   & 62.6 &
    ---                                                                                  \\
    \bottomrule
  \end{tabular}
\end{table}

%% file: sections/90_appendix_ssamba.tex
\section{SSAMBA Configuration and Inference Profiling}
\label{sec:appendix-ssamba}

This appendix collects the exploratory SSAMBA \cite{shams2024ssamba} experiments.
We record the patch-shape sweep, the inference profile against BEATs, and the full model chain.

We first experimented with reshaping the patch size on SSAMBA.
The default $16\times 16$ patch grid is inherited from ViT \cite{dosovitskiy2021vit} and is modified using a bidirectional Mamba scan.
We examined several configurations (Table~\ref{tab:ssamba-patch-shapes}), favoring tokens that capture more of the spectral information per time step.
The default tokenization uses only 1.5k tokens over a 30\,s clip, while a $32\times1$ grid uses 12k over the same period.
To see the benefits of $O(n)$ scaling, we need roughly 6{,}000 tokens in a 30\,s clip.
Against BEATs, SSAMBA is $3-4\times$ slower per clip but uses $\sim 3 \times$ less GPU memory (Table~\ref{tab:ssamba-vs-beats-inference}).
Without mamba-ssm CUDA kernels, SSAMBA is 22--130$\times$ slower, so the fused CUDA kernel or ONNX operator is required for these models to be viable at all.
We implemented an ONNX custom operator to run the Mamba selective scan efficiently on CPU.

\input{sections/floats/90_tab_ssamba-patch-shapes}

\input{sections/floats/90_tab_ssamba-vs-beats-inference}

Table~\ref{tab:ssamba-chain} lists the SSAMBA models we trained; none reaches the baselines reported by BirdSet \cite{rauch2025birdset} on supervised ConvNeXt \cite{liu2022convnext} and EfficientNet \cite{tan2019efficientnet}.
Linear probe is also a known-bad evaluator for MAE features (Bird-MAE \cite{rauch2025birdmae} shows 13 vs 49 mAP gap between linear and prototypical probes).
Results are inconclusive at our scale of 55k clips, an order of magnitude smaller than other SSL runs in the literature.

\input{sections/floats/90_tab_ssamba-chain}

%% file: sections/floats/90_tab_ssamba-patch-shapes.tex
\begin{table}[h]
  \centering
  \caption{
  SSAMBA patch-shape sweep.
  Forward-pass latency on RTX 6000 (24GB) for 30s audio, varying patch grid (frequency $\times$ time).
  Lower frequency-resolution rectangular patches are 2-2.3$\times$ faster than the standard 16$\times$16 square patches at equivalent token counts.
  }
  \label{tab:ssamba-patch-shapes}
  \begin{tabular}{lrrrr}
    \toprule
                                   &                  & \multicolumn{2}{c}{ms / fwd} &                                           \\
    \cmidrule(lr){3-4}
    Config                         & Tokens           & Tiny                         & Small         & Notes                     \\
    \midrule
    coarse-128$\times$24           & 125              & 28.3                         & 28.4          & overhead floor            \\
    fullband-128$\times$2          & 1{,}500          & 28.3                         & 28.3          & best at $\sim$1.5k tokens \\
    \textbf{baseline-16$\times$16} & \textbf{1{,}496} & \textbf{65.4}                & \textbf{65.8} &
    \textbf{SSAMBA default}                                                                                                \\
    framelevel-128$\times$1        & 3{,}000          & 29.2                         & 49.3          & 10\,ms temporal res       \\
    halfframe-64$\times$1          & 6{,}000          & 43.1                         & 90.3          & enters linear regime      \\
    quarterframe-32$\times$1       & 12{,}000         & 84.4                         & 182.3         & peak throughput           \\
    eighthframe-8$\times$1         & 48{,}000         & 490.5                        & 1{,}078       & superlinear (L2 spill)    \\
    \bottomrule
  \end{tabular}
\end{table}

%% file: sections/floats/90_tab_ssamba-vs-beats-inference.tex
\begin{table}[h]
  \centering
  \caption{
    Inference profile: SSAMBA-HF-small (26M) vs BEATs (90M) on V100-PCIE-16GB, 5s audio @ 16kHz, mamba-ssm CUDA kernels enabled.
  }
  \label{tab:ssamba-vs-beats-inference}
  \begin{tabular}{rrrrrr}
    \toprule
          & \multicolumn{2}{c}{Latency (ms)} & \multicolumn{2}{c}{GPU memory (MB)} &                                \\
    \cmidrule(lr){2-3} \cmidrule(lr){4-5}
    Batch & SSAMBA                           & BEATs                               & SSAMBA & BEATs   & Ratio (lat) \\
    \midrule
    1     & 30                               & 8                                   & 132    & 242     & 3.8$\times$ \\
    4     & 42                               & 13                                  & 205    & 496     & 3.3$\times$ \\
    16    & 150                              & 44                                  & 494    & 1{,}459 & 3.4$\times$ \\
    32    & 292                              & 92                                  & 881    & 2{,}759 & 3.2$\times$ \\
    \bottomrule
  \end{tabular}
\end{table}

%% file: sections/floats/90_tab_ssamba-chain.tex
\begin{table}[h]
  \centering
  \caption{SSAMBA + SSL experiments. cmAP on BirdSet HSN unless otherwise noted.}
  \label{tab:ssamba-chain}
  \begin{tabular}{llr}
    \toprule
    Setup                                          & Notes                             & cmAP \\
    \midrule
    SSAMBA-Tiny scratch                            & Random init                       &
    0.086                                                                                     \\
    SSAMBA-Tiny + AudioSet pretrain (16$\times$16) & Full transfer       &
    \textbf{0.179}                                                                            \\
    SSAMBA-Tiny + AudioSet pretrain (128$\times$1) & Mamba-only transfer &
    0.172                                                                                     \\
    SSAMBA-Tiny + SED + focal loss                 & Attention pool, 128$\times$1      &
    0.164                                                                                     \\
    SSAMBA-Small + SED + focal loss                & 25M params, no gain               &
    0.165                                                                                     \\
    SSAMBA-Tiny SSL (BirdCLEF 55K)                 & Linear probe          &
    0.009                                                                                     \\
    \midrule
    \multicolumn{3}{l}{\textit{References}}                                                   \\
    ConvNeXt (BirdSet LT)                          & Supervised pretrain               &
    0.47                                                                                      \\
    EfficientNet (BirdSet LT)                      & Supervised pretrain               &
    0.35                                                                                      \\
    Bird-MAE ViT-L (1.6M)                          & MAE + prototypical probe          &
    0.55                                                                                      \\
    \bottomrule
  \end{tabular}
\end{table}